\documentclass[twocolumn,prl,amsmath,amssymb,superscriptaddress,showpacs,floatfix,preprintnumbers]{revtex4}
\usepackage{graphicx}

\begin{document}
\title{Energy-Dependent Enhancement of the Electron-Coupling Spectrum \\ of the Underdoped Bi$_2$Sr$_2$CaCu$_2$O$_{8+\delta}$ Superconductor}
\author{H. Anzai}
\author{A. Ino}
\author{T. Kamo}
\author{T. Fujita}
\affiliation{Graduate School of Science, Hiroshima University, Higashi-Hiroshima 739-8526, Japan}

\author{M. Arita}
\author{H. Namatame}
\affiliation{Hiroshima Synchrotron Radiation Center, Hiroshima University, Higashi-Hiroshima 739-0046, Japan}

\author{M.~Taniguchi}
\affiliation{Graduate School of Science, Hiroshima University, Higashi-Hiroshima 739-8526, Japan}
\affiliation{Hiroshima Synchrotron Radiation Center, Hiroshima University, Higashi-Hiroshima 739-0046, Japan}

\author{\\ A. Fujimori}
\affiliation{Department of Physics, University of Tokyo, Tokyo 113-0033, Japan}

\author{Z.-X. Shen}
\affiliation{Department of Physics, Applied Physics and Stanford Synchrotron Radiation Laboratory, Stanford University, Stanford, California 94305, USA}

\author{M. Ishikado}
\altaffiliation{Present address: Japan Atomic Energy Agency, Tokai, Ibaraki 319-1195, Japan}
\affiliation{Department of Physics, University of Tokyo, Tokyo 113-0033, Japan}

\author{S. Uchida}
\affiliation{Department of Physics, University of Tokyo, Tokyo 113-0033, Japan}

\date{\today}

\begin{abstract}
We have determined the electron-coupling spectrum of superconducting Bi$_2$Sr$_2$CaCu$_2$O$_{8+\delta}$ from high-resolution angle-resolved photoemission spectra by two deconvolution-free robust methods. As hole concentration decreases, the coupling spectral weight at low energies $\lesssim$15 meV shows a twofold and nearly band-independent enhancement, while that around $\sim$65 meV increases moderately, and that in $\gtrsim$130 meV decreases leading to a crossover of dominant coupling excitation between them. Our results suggest the competition among multiple screening effects, and provide important clues to the source of sufficiently strong low-energy coupling, $\lambda_\mathrm{LE} \approx 1$, in an underdoped system.
\end{abstract} 
\pacs{74.72.-h, 71.18.+y, 74.25.Jb, 79.60.-i}
\maketitle


The coupling of electrons with other excitations plays an essential part in possible pairing mechanisms of superconductivity, and it concomitantly makes an electron appear as a slower and heavier quasiparticle. In reality, the electron is coupled with multiple kinds of excitations of various frequencies. Hence, the energy resolved data on the electron coupling provide important clues to the pairing glue. For high-$T_\mathrm{c}$ cuprates, it is believed that the strong electron correlation comes from the proximity to the Mott insulating phase \cite{Tokura1993PRL,ImadaFujimoriTokura1998RMP}. However, the behavior of the group velocity of renormalized quasiparticle is intriguing. With decreasing hole concentration, the decrease in velocity does not occur on the energy scale of the Mott transition \cite{XJZhou2003Nature,ALanzara2001Nature,PDJohnson2001PRL}, but at far lower energies, $<$40 meV, as reported recently \cite{IMVishik2010PRL,SESebastian2009Preprint}. The mechanism of this nontrivial mass-enhancement remains far from clear. In order to pin down the source of the coupling strength and the pairing attraction, we have to unravel the multilevel renormalization effects. Therefore, a thorough investigation of the energy, doping and band dependences of electron coupling is required.

Angle-resolved photoemission spectroscopy (ARPES) is an excellent tool for studying the interaction from the electron side \cite{IMVishik2010PRL,XJZhou2005PRL,WMeevasana2006PRL,TYamasaki2007PRB,KIshizaka2008PRB,WZhang2008PRL,JDRameau2009PRB,NCPlumb2010PRL}. Extracting the coupling information from subtle features in ARPES data has been attempted by the maximum-entropy method \cite{XJZhou2005PRL,WMeevasana2006PRL} and Richardson-Lucy method \cite{JDRameau2009PRB,NCPlumb2010PRL}. However, such deconvolution is a severe integral inversion problem, and possibly sensitive to statistical noise \cite{TVallaComment}. Here, we have developed two deconvolution-free robust methods for determining the ``effective'' coupling spectrum from high-resolution low-temperature ARPES spectra, noting the causal nature of the mass-enhancement factor defined as $\lambda(\omega) = - (d/d\omega) \Sigma(\omega)$, where $\Sigma(\omega)$ is electron self-energy \cite{GDMahan}.

In this paper, we report a systematic low-energy ARPES study of the electron-coupling spectrum of superconducting bilayer cuprates, Bi$_2$Sr$_2$CaCu$_2$O$_{8+\delta}$. Quantifying the impacts of three coupling features on the quasiparticle mass, we show that their contrasting evolutions with hole concentration cause a change in the dominant coupling excitation to occur. We propose possible scenarios for the mass enhancement with underdoping.

High-quality single crystals of Bi$_2$Sr$_2$CaCu$_2$O$_{8+\delta}$ were prepared by the traveling-solvent floating-zone method and a post annealing procedure. The hole concentration $p$ has been deduced from $T_c$ using a phenomenological relation, $T_c / T_c^\mathrm{max} = 1 - 82.6 (p - 0.16)^2$, where $T_c^\mathrm{max} = 91$ K \cite{1991MRPreslandPhysicaC}. Hereafter we label the samples by the doping level, i.e., underdoped (UD), optimally-doped (OP) or overdoped (OD), combined with the value of $T_c$. The ARPES spectra were collected at BL-9A of the Hiroshima Synchrotron Radiation Center using a Scienta R4000 electron analyzer. Instrumental energy and momentum resolution was 5 meV and 0.004 \AA$^{-1}$. The samples were cleaved \textit{in situ}, and kept under an ultrahigh vacuum (pressure under $5\times 10^{-11}$ Torr) at $T = 9$ K during the measurements.

\begin{figure}
\includegraphics[width=0.48\textwidth,bb=0 0 648 792]{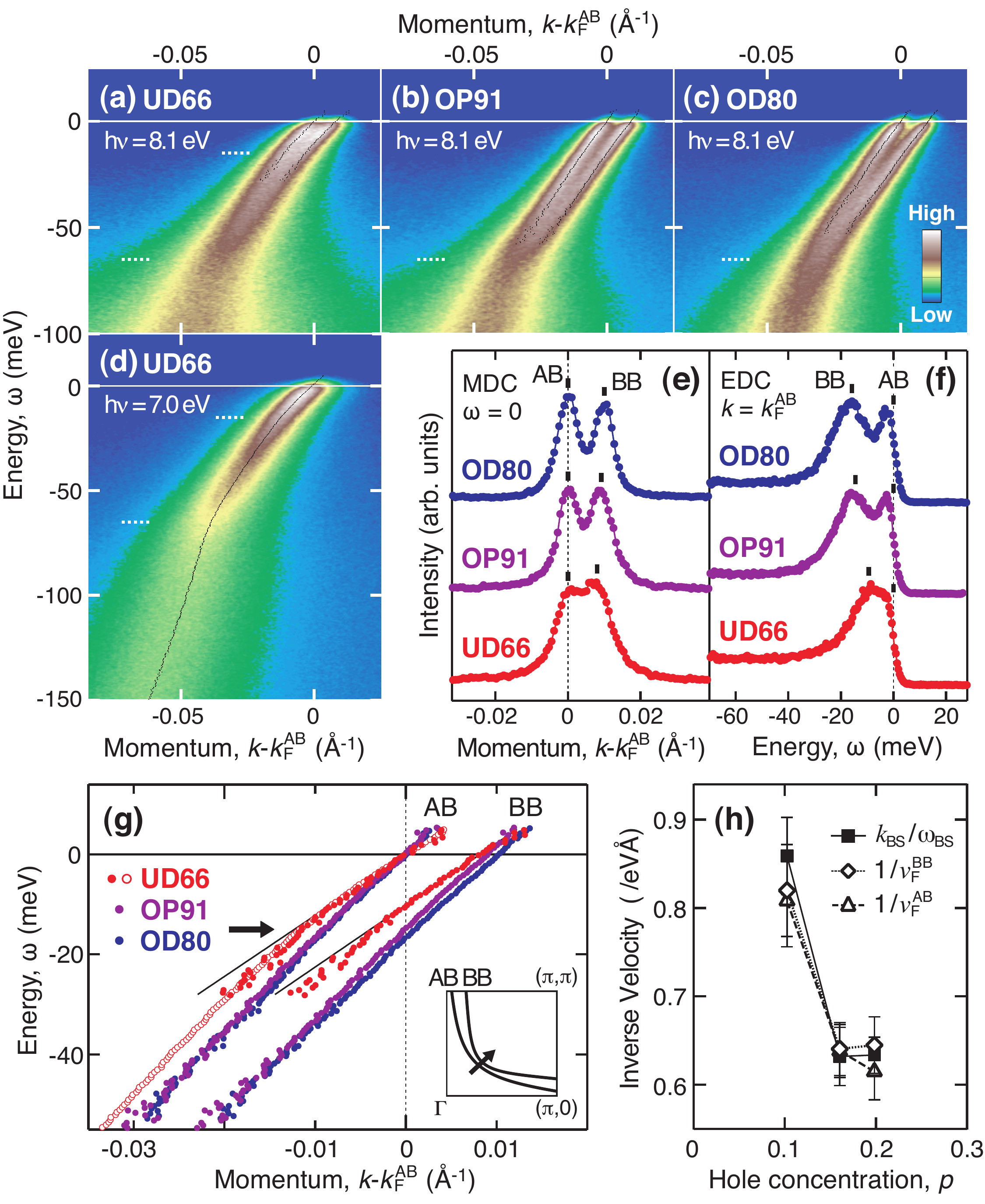}
\caption{(a)-(d) Energy-momentum plots of ARPES spectra along the nodal direction of Bi$_2$Sr$_2$CaCu$_2$O$_{8+\delta}$ for UD66 (underdoped, $T_\mathrm{c}=66$ K), OP91 (optimally-doped, $T_\mathrm{c}=91$ K), and OD80 (overdoped, $T_\mathrm{c}=80$ K) samples. (e) Momentum distribution curves (MDCs) at $\omega=0$ for $h\nu=8.1$ eV. (f) Energy distribution curves (EDCs) at $k=k_\mathrm{F}^\mathrm{AB}$ for $h\nu=8.1$ eV. (g) Quasiparticle dispersions determined by MDC fitting for UD66 (red), OP91 (purple), and OD80 (blue). Open and filled circles denote the results for $h\nu = 7.0$ and $8.1$ eV, respectively. (f) Inverse Fermi velocities of BB and AB, $1/v_\mathrm{F}^\mathrm{BB}$ (diamonds) and $1/v_\mathrm{F}^\mathrm{AB}$ (triangles), and momentum-to-energy ratio of bilayer splitting width, $k_\mathrm{BS} / \omega_\mathrm{BS}$ (filled squares).}
\label{spectra}
\end{figure}

Figures~\ref{spectra}(a)-(d) show the low-energy region of ARPES spectra. Despite a difficulty in controlling surface quality, a tiny nodal bilayer splitting was resolved even for the UD samples, as demonstrated in Figs.~\ref{spectra}(e) and (f). The full momentum width at half maximum, 0.009 \AA$^{-1}$, for UD66 is narrower than the previous studies \cite{KIshizaka2008PRB,IMVishik2010PRL}. We found that the spectral intensity ratio between the bilayer-split bands drastically changes with photon energy $h\nu$. We adopted $h\nu=8.1$ and 7.0 eV for simultaneous observation of the bonding band (BB) and the antibonding band (AB) in Figs.~\ref{spectra}(a)-(c), and for selective observation of AB in Fig.~\ref{spectra}(d), respectively.

The quasiparticle group velocity is given by the slope of dispersion, $v_\mathrm{g}(\omega) = d\omega/dk$. As hole concentration decreases, the splitting narrows more rapidly in energy than in momentum as shown in Figs.~\ref{spectra}(e), \ref{spectra}(f), and \ref{spectra}(h), providing a clear evidence for the decrease in Fermi velocity. The narrowing tendency with underdoping is consistent with what is expected from the decrease in out-of-plane conductivity \cite{XHChen}. In Figs.~\ref{spectra}(a)-(d), a dispersion kink is consistently observed at $|\omega| \sim 65$ meV \cite{ALanzara2001Nature,PDJohnson2001PRL}. Moreover, the dispersions at low energies, $|\omega| \lesssim 15$ meV, are substantially curved for UD66, whereas they seem relatively straight for OP91 and OD80. We can rule out the effect of transition-matrix elements, because the dispersions determined with different photon energies are identical as shown in Fig.~\ref{spectra}(g). Comparing the Fermi velocities of BB and AB in Fig.~\ref{spectra}(h), we find that BB exhibits an effective-mass enhancement similar to AB with underdoping \cite{IMVishik2010PRL}. These results indicate that the low-energy interaction is nearly independent of the bilayer bands.

\begin{figure*}
\includegraphics[width=\textwidth,bb=0 0 920 491]{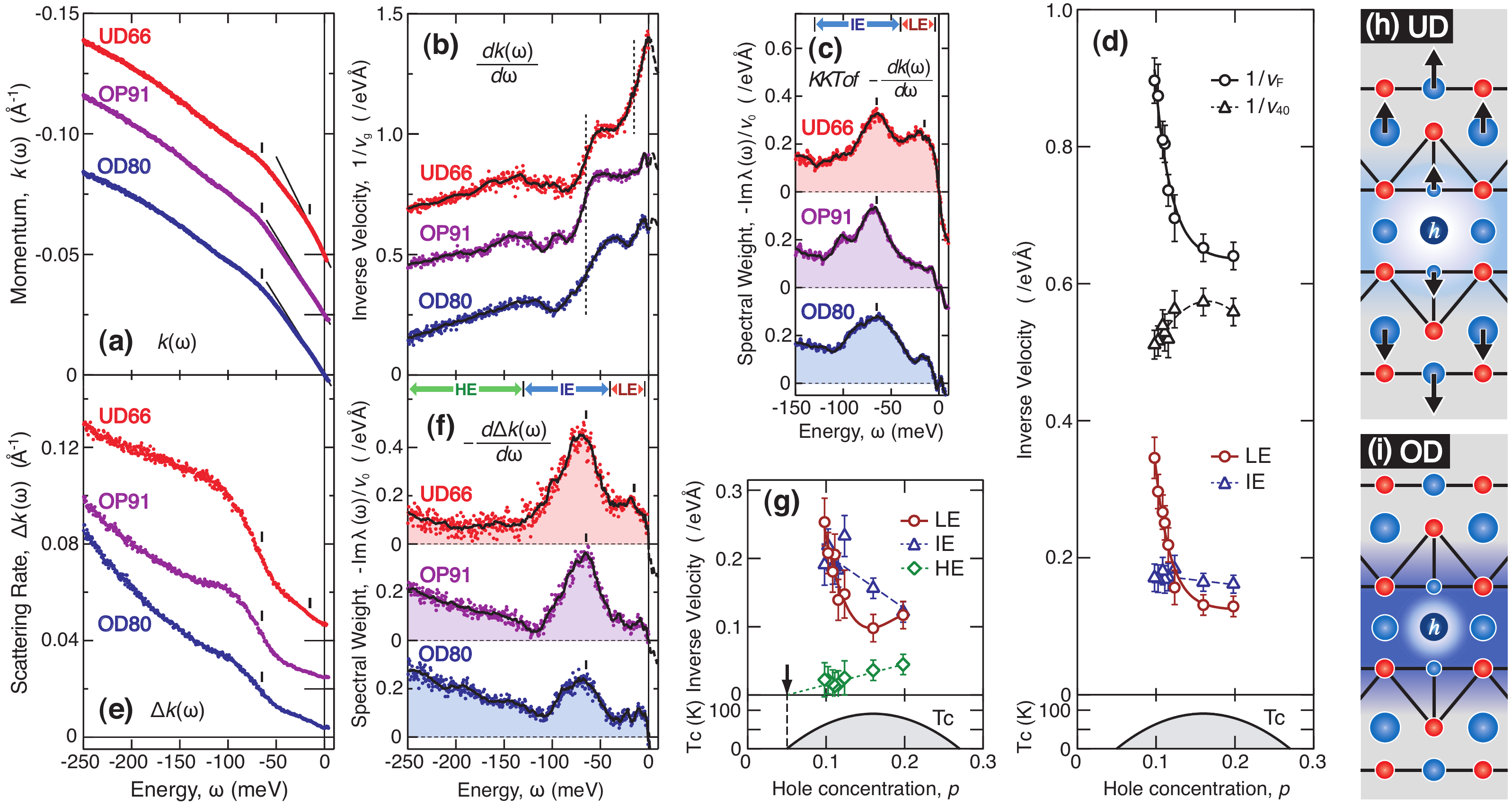}
\caption{(a) Quasiparticle dispersion determined from MDC-peak position, $k(\omega) = [\omega - \mathrm{Re}\Sigma(\omega)]/v_0$. (b) Inverse group velocity, $1/v_g(\omega) = [1 + \mathrm{Re}\lambda(\omega)]/v_0$, determined from $(d/d\omega) k(\omega)$. (c) Effective coupling spectra, $-\mathrm{Im}\lambda(\omega)/v_0$, deduced from the Kramers-Kronig transform (KKT) of $-(d/d\omega) k(\omega)$ \cite{extrapolation}. (d) Doping dependence of inverse group velocities at Fermi level, $1/v_F$ (black circles), and at 40 meV, $1/v_{40}$ (black triangles), and partial coupling constants, $\lambda/v_0$ (color), deduced from the KKT of $(d/d\omega) k(\omega)$. (e) Quasiparticle scattering rate determined from MDC-peak natural half width excluding instrumental resolution, $\Delta k(\omega) = - \mathrm{Im}\Sigma(\omega)/v_0$. (f) Effective coupling spectrum, $-\mathrm{Im}\lambda(\omega)/v_0$, directly determined from $-(d/d\omega) \Delta k(\omega)$. (g) Doping dependence of partial coupling constants, $\lambda/v_0$, deduced from $(d/d\omega)\Delta k(\omega)$. (h) Strong electron-phonon coupling at low hole concentration. (i) Weak electron-phonon coupling at high hole concentration. The low-energy (LE, red circles), intermediate-energy (IE, blue triangles), and high-energy (HE, green diamonds) parts are defined as $4<|\omega|<40$ meV, $40<|\omega|<130$ meV, and $130<|\omega|<250$ meV, respectively. The differential coefficient at $\omega$ has been evaluated within an energy window of $\omega - W(\omega) \le \omega \le \omega + W(\omega)$, where $W(\omega) = 5.5 + 0.15 |\omega|$ meV, from simple difference between both ends of the window (color dots) and by least-squares linear regression method (black curves). The data are offset for clarity in (a), (b) and (e).}
\label{dopingdependence}
\end{figure*}

\begin{figure}
\includegraphics[width=0.42\textwidth,bb=0 0 573 348]{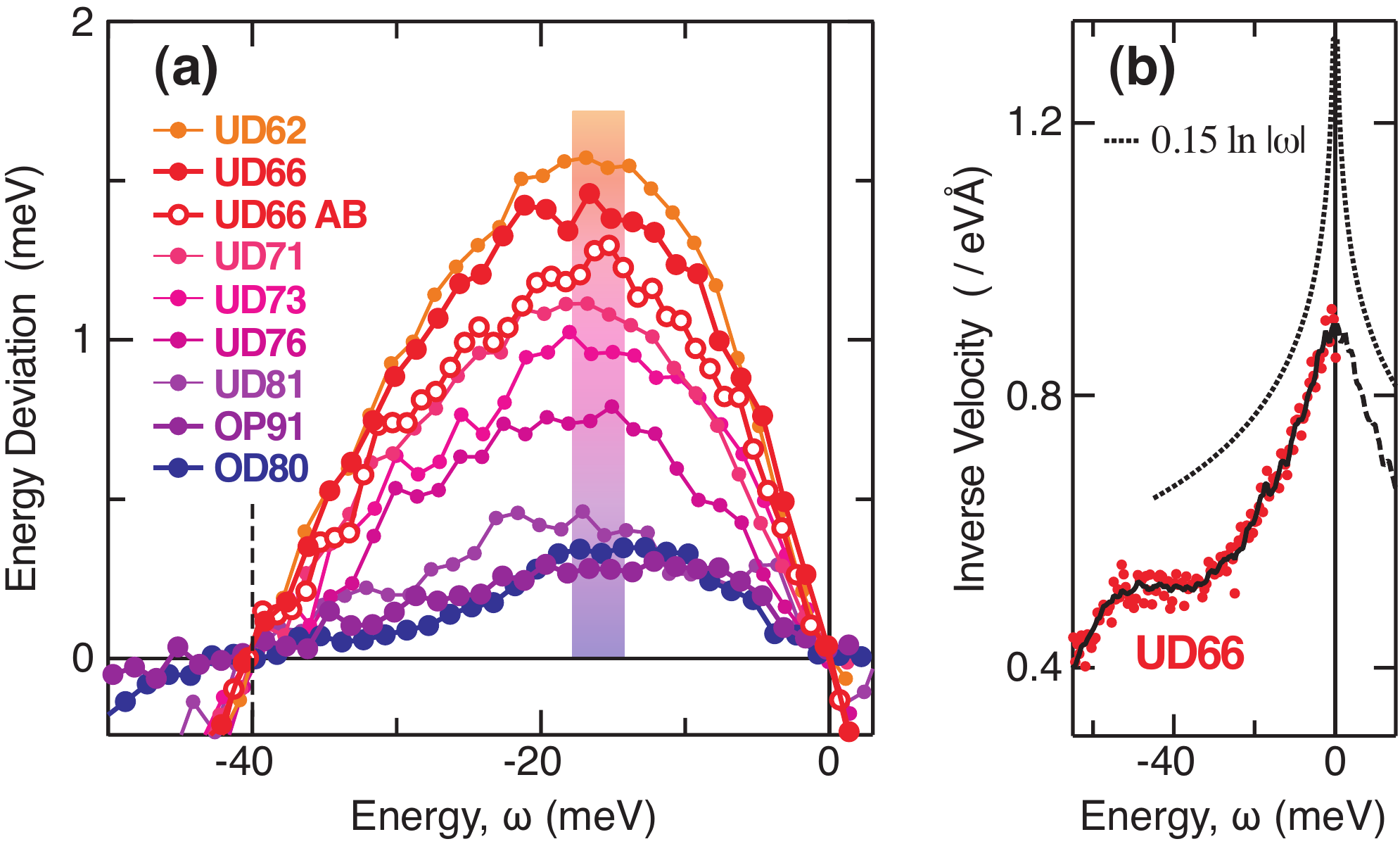}
\caption{Dispersion anomaly at low energies. (a) Energy deviation from the straight line which intersects with the experimental dispersion at $\omega = -40$ and 0 meV. Filled and open marks denote the result for $h\nu=8.1$ and 7.0 eV, respectively. (b) Inverse group velocity $1/v_g(\omega)$ for UD66, compared with a logarithmic function, $0.15 \ln |\omega|$ (dotted curve).}
\label{deviation}
\end{figure}

Thus, we have determined the quasiparticle dispersion and scattering rate over a wide energy range by imposing $\omega$-independent bilayer-splitting parameters on the fitting analysis of momentum distribution curves (MDCs). Figures~\ref{dopingdependence}(a) and \ref{dopingdependence}(e) show the peak position $k(\omega)$ and natural half-width $\Delta k(\omega)$, respectively. Approximating bare-electron velocity $v_0$ by a constant, we obtain the forms, $k(\omega) = [\omega - \mathrm{Re}\Sigma(\omega)]/v_0$ and $\Delta k(\omega) = -\mathrm{Im}\Sigma(\omega)/v_0$. Figure~\ref{deviation}(a) shows that a small dispersion kink at $\sim$10 meV \cite{WZhang2008PRL,JDRameau2009PRB,NCPlumb2010PRL} evolves with underdoping into a large and broad feature around $\sim$15 meV. The difference between the inverse group velocities at the Fermi level and at 40 meV shown in Fig.~\ref{dopingdependence}(d) indicates that the coupling with low-energy excitations is abruptly enhanced upon entering the underdoped region.

The real and imaginary parts of mass enhancement, $1+\lambda(\omega) = v_0/v_\mathrm{g}(\omega)$, are directly deduced from the energy derivatives of $k(\omega)$ and $\Delta k(\omega)$, respectively, by the forms, 
\begin{eqnarray}
\frac{dk(\omega)}{d\omega} &=& \frac{1 + \mathrm{Re}\lambda(\omega)}{v_0} = \frac{1}{v_\mathrm{g}(\omega)},\nonumber \\
\frac{d\Delta k(\omega)}{d\omega} &=& \frac{\mathrm{Im}\lambda(\omega)}{v_0}.\nonumber 
\end{eqnarray}
The differential scattering rate, $\mathrm{Im}\lambda(\omega)$, at $T=0$ represents a kind of coupling spectral function, which includes the effects of $k$ and $\omega$ dependences of electronic spectral function. Note that the $\omega$ dependence of $1+\lambda(\omega)$ is irrelevant to the uncertainty of $v_0$ unlike that of $\mathrm{Re}\Sigma(\omega)$.

The energy dependence of mass enhancement $\mathrm{Re}\lambda(\omega)$ is presented as $1/v_\mathrm{g}(\omega)$ in Fig.~\ref{dopingdependence}(b). A steplike mass increase at $\sim$ 65 meV is the typical behavior of electron coupling with a certain bosonic mode, and in good agreement with the results of optical studies \cite{optical}. By contrast, a cusplike mass enhancement at $\omega=0$ for UD66 appears to have no saturation of slope, as shown in Fig.~\ref{deviation}(b). This indicates that the onset of coupling spectral weight is quite close to $\omega=0$, and may cause virtually singular behavior of the quasiparticles.

The effective coupling spectra $\mathrm{Im}\lambda(\omega)$ have been deduced from the Kramers-Kronig transform (KKT) of $(d/d\omega) k(\omega)$ and directly from $(d/d\omega) \Delta k(\omega)$, as shown in Figs.~\ref{dopingdependence}(c) and \ref{dopingdependence}(f), respectively \cite{extrapolation}. Furthermore, dividing the integral, 
$$\lambda = \mathrm{Re}\lambda(0) = \frac{2}{\pi}\int^{\infty}_{0}\frac{\mathrm{Im}\lambda(\omega)}{\omega}d\omega,$$
into three energy parts, we have quantified the partial coupling constants as shown in Figs.~\ref{dopingdependence}(d) and \ref{dopingdependence}(g). Although the experimental accuracy is better for the peak position $k(\omega)$ than for the peak width $\Delta k(\omega)$, the derivation is more direct in Fig.~\ref{dopingdependence}(f), and the slope of the spectral background in Fig.~\ref{dopingdependence}(c) remains uncertain due to the extrapolation required for KKT \cite{extrapolation}. To this extent, the results from $k(\omega)$ and $\Delta k(\omega)$ are consistent. As hole concentration decreases, whereas the intermediate-energy (IE) part around the 65-meV peak shows moderate increase \cite{ALanzara2001Nature,PDJohnson2001PRL,optical}, the low-energy (LE) part abruptly shows twofold enhancement with a curve quite similar to that of $1/v_\mathrm{F}$. This behavior is manifest not only in the curvature of the dispersion $k(\omega)$ in Figs.~\ref{dopingdependence}(a) and \ref{deviation}(a), but also in the slope of the scattering rate $\Delta k(\omega)$ in Fig.~\ref{dopingdependence}(e), and is consistent with the $\omega$-linear term of scattering rate deduced from tunneling spectra \cite{JWAlldredge2008NP}. Furthermore, Figs.~\ref{dopingdependence}(d) and \ref{dopingdependence}(g) show that the impact of the LE part exceeds that of the IE part on the underdoped side, implying a crossover of the dominant coupling excitations from $\sim$65 meV to $\lesssim$15 meV. Assuming $v_0\approx 4$ eV{\AA} based on the local-density-approximation calculation \cite{RSMarkiewicz2005PRB}, sufficiently strong coupling of order unity, $\lambda^\mathrm{LE}\approx 1$, is realized only with the low-energy excitations.


The energy-dependent enhancement of the coupling spectrum with underdoping is likely related to high-energy electron-electron interaction \cite{WMeevasana2006PRL}. In Fig.~\ref{dopingdependence}(f), one finds that the substantial coupling weight linear in $\omega$ extends over $|\omega|>130$ meV beyond the phonon cutoff. Moreover, Fig.~\ref{dopingdependence}(g) suggests that the high-energy (HE) part decreases to zero towards the superconductor-to-insulator transition point. Therefore, this part should be ascribed to the electron-electron interaction expressed as $\mathrm{Im}\Sigma^\mathrm{e-e} \propto \omega^2$. With sufficient hole concentration, the bare Coulomb potential is quickly screened by these high-frequency electronic excitations, and thereby the retarded response of low-frequency excitations is suppressed. Such suppression becomes more drastic as the frequency of the coupling excitation decreases. Thus, the contrasting behaviors of the LE, IE, and HE parts are consistently interpreted as the competing effect in screening the Coulomb potential.

A candidate for the origin of the large effective mass in the underdoped system is the bare strong coupling with low-frequency optical phonons as illustrated in Fig.~\ref{dopingdependence}(h). Rameau \textit{et al.~}assigned the small low-energy kink of optimally-doped Bi$_2$Sr$_2$CaCu$_2$O$_{8+\delta}$ to the $c$-axis phonon involving the vibration of Bi and Sr atoms \cite{JDRameau2009PRB,NNKovaleva2004PRB}. At the node of $d$-wave gap, the low-energy scattering is only possible without in-plane momentum transfer. The phonons of out-of-plane momentum can provide such forward scatterings, and thus are compatible with the $d$-wave superconductivity also as a pairing interaction \cite{MLKulic}.

In addition, the effective mass may also be enhanced near the quantum critical point between the metallic and insulating phases, as proposed from the quantum oscillations \cite{SESebastian2009Preprint}. The instability toward some competing orders of charge or spin induces the divergent critical fluctuations \cite{SAKivelson2003RMP,CMVarma2002PhysRep,THanaguri2004Nature}, even though the optical phonons make no contribution to the mass divergence. Phenomenologically, the LE parts of $\mathrm{Re}\lambda(\omega)$ and $\mathrm{Im}\lambda(\omega)$ for UD66 resemble the marginal-Fermi-liquid form, $\lambda(\omega) \propto \ln\left|\omega_\mathrm{c}/\omega\right| -1 - i \pi/2$ for $\omega < 0$ where $\omega_\mathrm{c}$ is cutoff energy, as shown in Figs.~\ref{deviation}(b), \ref{dopingdependence}(b), \ref{dopingdependence}(c) and \ref{dopingdependence}(f) \cite{CMVarma2002PhysRep}.

Finally, we note the effect of elastic forward impurity scattering. Tunneling spectroscopy measurements have revealed that nanoscale areas, where the coherence peak is absent, emerges for $p<0.13$ \cite{KMcElroy2005PRL,KMcElroy2005Science}. The local depletion of quasiparticles implies breakdown of the static screening, thus giving rise to the static-potential inhomogeneity that originates from out-of-plane distant impurities \cite{KMcElroy2005Science}, and would result in a dramatic increase in elastic forward scattering. Given the conelike dispersion around the node of the $d$-wave gap, the elastic scattering rate may have quasilinear energy dependence \cite{TYamasaki2007PRB,KIshizaka2008PRB,TDahm2005PRB}. In that case, the energy and mass of nodal quasiparticle may also be renormalized by the second-order forward scattering process.


In conclusion, our ARPES study has revealed the nearly band-independent and contrastingly energy-dependent evolution of the electron-coupling spectrum with hole concentration. In underdoped Bi$_2$Sr$_2$CaCu$_2$O$_{8+\delta}$, the strong coupling weight ($\lambda_\mathrm{LE} \approx 1$) is distributed around $\sim$15 meV with an onset at $\omega \simeq 0$. As hole concentration decreases, the LE part shows twofold rapid enhancement, the IE part increases moderately, and, by contrast, the HE part decreases to zero towards the superconductor-to-insulator transition point. This behavior suggests the competition among multiple screening effects as a possible origin of the mass enhancement. In terms of the dominant coupling excitation, a crossover from $\sim$65 meV to $\lesssim$15 meV occurs upon underdoping, whereas the electron-electron part in $\gtrsim$130 meV increase in presence with overdoping. The balance among these multiple interactions provides a new perspective on the phase diagram of cuprates. In particular, whether the low-energy interaction is pair-breaking or pair-binding would be an important subject of future study.


This work was supported by KAKENHI (20740199). H.~A.~acknowledges financial support from JSPS. The ARPES experiments were performed under the approval of HRSC (Proposal No.~07-A-2).


\begin{thebibliography}{99}
\bibitem{extrapolation}
Symmetrization with respect to $\omega =0$ and constant extrapolation for $|\omega|\gtrsim 0.28$ eV were applied before KKT.
\end{thebibliography}
\end{document}